\documentclass[12pt]{article}
\usepackage{cite}
\usepackage[utf8]{inputenc}
\usepackage{amsmath, amssymb}
\usepackage{geometry}
\usepackage{hyperref}
\usepackage{tikz}
\usepackage{young}
\usepackage{youngtab}
\usepackage{ytableau}
\usepackage{ulem}
\usepackage{verbatim}
\usetikzlibrary{arrows.meta, positioning}
\newcommand{\bea}{\begin{eqnarray}}
\newcommand{\eea}{\end{eqnarray}}

\newcommand{\fund}{\raisebox{-0.4ex}{$\text{\yng(1)}$}}
\newcommand{\antifund}{\raisebox{-0.6ex}{$\overline{\text{\yng(1)}}$}}

\begin{document}
\newcommand\mytitle{The Potency of Nilpotence}
\newcommand\mypreprint{UCI-TR-2025-24}
\begin{titlepage}
	\begin{flushright}
		\mypreprint
	\end{flushright}
	
	\vspace*{2cm}
	
	\begin{center}
		{\Large\sffamily\bfseries\mytitle}
		
		\vspace{1cm}
		
		\renewcommand*{\thefootnote}{\fnsymbol{footnote}}
		
		\textbf{%
			Eric Bryan\footnote{ebryan2@uci.edu}, 
			Arvind Rajaraman\footnote{arajaram@uci.edu}, and
			Yuri~Shirman\footnote{yshirman@uci.edu},
		}
		\\[8mm]
		\textit{\small
			~Department of Physics and Astronomy\\ University of California\\ Irvine, CA 92697-4575, USA
		}
	
	\end{center}

	\vspace*{1cm}
	\begin{abstract}
	The dynamics of $\mathcal{N}=1$  SUSY gauge theories with matter in adjoint and fundamental representations and the superpotentials given by Arnold's ADE singularities has been extensively studied in the literature. It was also conjectured that supersymmetric models with $W_{A_k}$, $W_{D_{k+2}}$ and $W_{E_7}$ superpotentials possess a dual description. In this paper we revisit the analysis of the moduli space of $A_k$ and $D_{k+2}$ models by considering the duality along nilpotent directions on the moduli space. While our analysis provides additional evidence for the duality conjecture in $W_{A_k}$ models, we show that  the duality conjecture  fails for the $W_{D_{k+2}}$ models. 
	\end{abstract}
	\vspace*{1cm}
\end{titlepage}

\setcounter{footnote}{0}
\section{Introduction}

One of the most striking features of $\mathcal{N}=1$ supersymmetric gauge theories is Seiberg duality~\cite{Seiberg:1994pq}, which posits an infrared (IR) equivalence between seemingly different ultraviolet (UV) gauge theories.  The initial examples of dualities in $SU(N)$ theories with matter in a fundamental representation were quickly expanded to more general theories, and even to chains of dualities.

The duality conjectures have been tested and extended in numerous settings, particularly for theories with simple superpotentials and gauge groups. Since at least one of the two dual descriptions is typically strongly coupled in the IR, the duality is usually verified by indirect checks. Firstly, the global symmetries of the two theories must agree. Secondly, the 't Hooft 
anomalies must match for the two theories. These are very restrictive conditions, and 
the successful matching of the 't Hooft  anomalies is taken as strong evidence for the  duality. 

Further tests of the dualities involve  dynamics on the moduli space or the deformation of the theory by new superpotential terms. Often, these tests are complementary -- turning on moduli VEVs on the electric side corresponds to superpotential deformations on the magnetic side and vice versa. Typically, turning on gauge invariant operators on the electric side of duality results in gauge symmetry breaking. On the magnetic side, this usually corresponds to superpotential deformations which may result in new Yukawa interactions, mass terms for some matter fields, and gauge symmetry breaking. For the duality to be valid, the two low energy theories must remain dual. This check is especially non-trivial along flat directions where one can show that the two low energy theories flow to the same IR fixed point.

A comprehensive study of dynamics on the moduli space has been done for some classes of models, for example, $SU(N)$ theories with matter in the fundamental representation \cite{Seiberg:1994pq}. However, due to the complexity of the moduli space, the analysis is far from complete in models with more general gauge groups, matter content, and the superpotentials. 

In this paper we will focus on   $\mathcal{N}=1$ dualities in  models with chiral superfields in both fundamental and adjoint representations and nontrivial superpotentials for the adjoints. While the exhaustive study of the dynamics on the moduli space of these theories is beyond the scope of this paper, we will extend the analysis to a new class of flat directions (we will refer to such flat directions as nilpotent) and show that new stringent tests of dualities can be performed along these directions. We expect such analysis to be useful in more general models with general matter representations and non-trivial superpotentials which allow the existence of nilpotent flat directions.

 The first  model of $\mathcal{N}=1$ duality  with adjoint superfields  was proposed by Kutasov~\cite{Kutasov:1995ve,Kutasov:1995np,Kutasov:1995ss}, where the electric description was based on an $SU(N)$ gauge group with {$F$ chiral flavors in the fundamental representation,} a single adjoint chiral superfield $X$ and a superpotential\footnote{We will ignore numerical coefficients throughout.} 
\begin{equation}
	W=\mathrm{Tr} X^{k+1}\,.
\end{equation}
 The $F$-flatness conditions coming from the superpotential  truncate the chiral ring of the theory and are an essential condition for the existence of the duality map.

Kutasov's conjecture was extended in ~\cite{Brodie:1996vx} to theories with {$F$ fundamental flavors,} two adjoints $X$ and $Y$, and a superpotential
\begin{equation}
	W=\mathrm{Tr} X^{k+1}+ \mathrm{Tr} XY^2\,.
\end{equation}
It was also shown~\cite{Brodie:1996vx,Intriligator:2003mi,Intriligator:2016sgx,Kutasov:2014yqa} that the superpotentials in these two classes of models correspond to $A_k$ and $D_{k+2}$ singularities of Arnold's singularity theory~\cite{Arnold1981}. 

However, already in the original paper \cite{Brodie:1996vx} it was realized that the duality conjecture in $D_{k+2}$ poses a puzzle. For any value of $k$, the theory and the dual have the same symmetries, and the global anomalies match across the theories. However, while the $F$-term equations lead to truncation of the chiral ring in models with odd $k$, the chiral ring is not truncated in models with even $k$. Since the truncation of the chiral ring is  a necessary condition for the existence of duality, this puts in question the validity of the duality conjecture of \cite{Brodie:1996vx}, at least when $k$ is even. While one may have more confidence in the duality for  odd $k$, there is still a  tension, which  arises from the fact that one can flow between even and odd $k$ theories either by deforming them  with tree level superpotential terms or by simply moving along the moduli space.

One possible resolution of this puzzle  could be that the chiral ring of even $k$ model is truncated by non-perturbative dynamics. Further research led to better understanding of the dynamics of ADE models~\cite{Intriligator:2016sgx}, however, no dynamical effects capable of truncating the chiral ring were found.

In this work we will take a different approach towards addressing the puzzle. In particular, we will carefully investigate a  class of directions  on the moduli space of the $A_k$ and $D_{k+2}$ models along which the VEVs of $X$ is nilpotent\footnote{When studying $D_{k+2}$ models we will limit our attention to $Y=0$ directions for simplicity.}, $X^k=0$. In the case of $A_k$ models our work will extend the results of~\cite{Aharony:1995ne} while in the case of $D_{k+2}$ models our analysis will be completely new. Our results will further strengthen the evidence for duality in $A_k$ models but  the tests of duality will fail in $D_{k+2}$ models even when $k$ is odd.

The outline of the paper is as follows. In  Section \ref{sec:review}, we review the conjectured dualities involving adjoints. In Section \ref{sec:newtests} we discuss the tests of duality in $A_k$ and $D_{k+2}$ models. After reviewing the existing duality tests in Section \ref{sec:existing} we discuss the dynamics along nilpotent flat directions in Section \ref{sec:newtests}. We then test the duality of $A_k$ models in Section \ref{sec:Aktests} finding additional evidence in support of duality conjecture. Finally, in Section \ref{sec:Dktests} we show that similar analysis of $D_{k+2}$ models does not yield consistent results. We close with a discussion.

\section{Review of $A_k$ and $D_{k+2}$ Dualities }
\label{sec:review}

\subsection{$A_k$ Dualities}
The electric description of $A_k$ models \cite{Kutasov:1995ve} is given in terms of an $SU(N)$ gauge group with an adjoint superfield $X$ and $F$ flavors in the fundamental representation. The matter content and symmetries of the model are given by Table \ref{tab:akElUV}.
\begin{table}[h!]
\centering
\def\arraystretch{1.5}
\begin{tabular}{||c | c | c | c| c| c|} 
 \hline
 Field & $SU(N)$  & $SU(F)$  & $SU({F})$ & $U(1)_B$ & $U(1)_R$\\ [0.5ex] 
 \hline\hline
 $ Q$ & \fund & \fund & 1&1&$1-\frac{2}{k+1}\frac{N}{F}$  \\ 
 \hline
 $\bar{Q}$ & \antifund  & 1 & \antifund&-1&$1-\frac{2}{k+1}\frac{N}{F}$ \\
 \hline
 ${X}$ & Adj & 1 & 1 &0&$\frac{2}{k+1}$\\  
 \hline
\end{tabular}
\caption{Matter content and global charges of the $A_k$ electric theory.}
\label{tab:akElUV}
\end{table}
The $U(1)_R$ charges were chosen so that the following superpotential is allowed by symmetries
\begin{align}
	\label{eq:Akelectric}
W_{el}= \mathrm{Tr}(X^{k+1})+\lambda \mathrm{Tr} X\,,
\end{align}
where we have included a Lagrange multiplier term $\lambda\mathrm{Tr}X$ in the superpotential to enforce tracelessness
of the adjoint.
The equation of motion ensures that $X$ is nilpotent up to an identity operator
\begin{equation}
    X^k=\lambda \mathbb{I}\,.
\end{equation}
We will be interested in the moduli space directions consistent with $\lambda=0$ and therefore we will drop the Lagrange multiplier terms in the following.

As was argued in \cite{Kutasov:1995ss}, this equation truncates the chiral ring of the theory and deforms the quantum moduli space. The low energy physics can then be described in terms of ``dressed" chiral  meson and baryon operators that belong to the truncated chiral ring. In particular, the meson operators in the chiral ring are of the form 
\begin{equation}
M_m=\bar{Q}X^{m-1}Q\,,~~~~ m={1,2,...,k}\,.
\end{equation}
The duality maps dressed mesons to the elementary mesons in the magnetic theory\footnote{The dressed baryons are mapped to dressed baryons of the magnetic description but we will not write them down explicitly since we will not turn them on in our discussion.}. The magnetic theory is described in terms of an $SU(kF-N)$ gauge theory with matter content provided in table \ref{tab:akMagUV}.
\begin{table}[h!]
\centering
\def\arraystretch{1.5}
\begin{tabular}{||c | c | c | c| c| c|} 
 \hline
 Field & $SU(kF-N)$  & $SU(F)$  & $SU({F})$ & $U(1)_B$ & $U(1)_R$\\ [0.5ex] 
 \hline\hline
 $ q$ & \fund & \antifund & 1&$\frac{N}{kF-N}$&$1-\frac{2}{k+1}\frac{kF-N}{F}$ \\ 
 \hline
 $\bar{q}$ & \antifund  & 1 & \fund &$-\frac{N}{kF-N}$&$1-\frac{2}{k+1}\frac{kF-N}{F}$\\
 \hline
 $\tilde{X}$ & Adj & 1 & 1&0&$\frac{2}{k+1}$ \\  
 \hline
 $M_m$ &1&\fund&\antifund&0&$2-\frac{4}{k+1}\frac{N}{F}+\frac{2}{k+1}(m-1)$\\
 \hline
\end{tabular}
\caption{Matter content and global charges of the $A_k$ magnetic theory.}
\label{tab:akMagUV}
\end{table}
and the superpotential
\begin{align}
   W_{mag}&= \mathrm{Tr}(\tilde{X}^{k+1})+\sum_{m=1}^k M_m \bar{q}\tilde{X}^{k-m} q 
\end{align}

\subsection{$D_{k+2}$ Dualities}
\label{sec:dkplus2}
The $D_{k+2}$ model \cite{Brodie:1996vx} can be obtained by extending the $A_k$ model with a second adjoint superfield, $Y$, and a new superpotential term. The electric description is based on an $SU(N)$ gauge group with
matter content given by Table \ref{tab:dkElUV}
\begin{table}[h!]
\centering
\def\arraystretch{1.5}
\begin{tabular}{||c | c | c | c| c| c|} 
 \hline
 Field & $SU(N)$  & $SU(F)$  & $SU({F})$ & $U(1)_B$ & $U(1)_R$\\ [0.5ex] 
 \hline\hline
 $ Q$ & \fund& \fund & 1 &$1$&$1-\frac{N}{F(k+1)}$\\ 
 \hline
 $\bar{Q}$ & \antifund  & 1 & \antifund & $-1$&$1-\frac{N}{F(k+1)}$ \\
 \hline
 ${X}$ & Adj & 1 & 1&0&$\frac{2}{k+1}$ \\  
 \hline
 ${Y}$ & Adj & 1 & 1 &0&$\frac{k}{k+1}$\\  
 \hline
\end{tabular}
\caption{Matter content and global charges of the $D_{k+2}$ electric theory.}
\label{tab:dkElUV}
\end{table}
and the superpotential\footnote{As mentioned earlier, we have dropped Lagrange multiplier terms for brevity.}
\begin{align}
W= \mathrm{Tr}\left(X^{k+1}\right)+ \mathrm{Tr}\left(XY^2\right)\,,
\end{align}
The equations of motion are given by
\begin{equation}
\begin{split}
	\label{eq:dkeom}
    & (k+1) {X}^{k}+ {Y}^{2}=0\\ 
  & XY+YX= 0
\end{split}
\end{equation}
 
Clearly, in this model $X$ is only nilpotent if $Y$ is. On the other hand, we can see from (\ref{eq:dkeom}) that $Y$ satisfies the equation
\begin{equation}
	Y^3=\left(-1\right)^{k} Y^3\,,
\end{equation}
implying that it is nilpotent for odd $k$ while remaining unconstrained for even $k$.
Thus the chiral ring is only truncated for odd values of $k$.  It has been conjectured in the literature that for even $k$ the chiral ring is truncated by non-perturbative dynamics, however no such dynamical effects have been found so far \cite{Intriligator:2016sgx}.

Assuming the truncation of the chiral ring, the mesonic directions of the quantum moduli space of the electric theory are parameterized by
 \begin{equation}
 	\label{eq:dkplustwomeson}
     M_{m,l}=\bar{Q}X^{m-1}Y^{l-1}Q\,,~~~~~m=1,2\dots,k\,;~~ l=1,2,3\,.
 \end{equation}
As usual, the composite mesons map into the elementary mesons of the magnetic description. The dual theory has gauge group $SU(3kF-N)$ and matter content is given by Table \ref{tab:dkMagUV}.
\begin{table}[h!]
\centering
\def\arraystretch{1.5}
\begin{tabular}{||c | c | c | c| c| c|} 
 \hline
 Field & $SU(3kF-N)$  & $SU(F)$  & $SU({F})$ & $U(1)_B$ & $U(1)_R$\\ [0.5ex] 
 \hline\hline
 $ q$ & \fund& \antifund & 1&$\frac{N}{3kF-N}$&$1-\frac{3kF-N}{F(k+1)}$ \\ 
 \hline
 $\bar{q}$ & \antifund  & 1 & \fund & $-\frac{N}{3kF-N}$ &$1-\frac{3kF-N}{F(k+1)}$\\
 \hline
 $\tilde{X}$ & Adj & 1 & 1&0&$\frac{2}{k+1}$ \\ 
 \hline
 $\tilde{Y}$ & Adj & 1 & 1&0&$\frac{k}{k+1}$ \\ 
 \hline
 $M_{m,l}$ &1&\fund&\antifund&0&$2-\frac{2}{k+1}\frac{N}{F}+\frac{2(m-1)+k(l-1)}{k+1}$\\
 \hline
\end{tabular}
\caption{Matter content and global charges of the $D_{k+2}$ magnetic theory.}
\label{tab:dkMagUV}
\end{table}
The superpotential of the magnetic dual is given by
\begin{equation}
	\label{eq:Dkplus2magneticW}
	W=\mathrm{Tr}\tilde{X}^{k+1}+\mathrm{Tr} \tilde{X}\tilde{Y}^2+\sum_m^k\sum_l^3 M_{m,l}\bar q \tilde{X}^{k-m}\tilde{Y}^{3-l}q\,,
\end{equation}

\section{ New Tests of Duality  Conjectures}
\label{sec:newtests}

\subsection{Existing Tests of the Dualities}
\label{sec:existing}
The
duality conjectures in $A_k$ and $D_{k+2}$ theories have been tested in a variety of ways.
The global symmetries and anomalies of the electric  and magnetic  description agree. The t'Hooft anomalies of the two theories have been shown to match (notably, for the $D_{k+2}$ theories, this agreement occurs for both odd and even $k$). Certain deformations of the superpotential, e.g. by mass terms, have also been considered, and the duality conjecture  continues to hold.

A different set of tests considers  the dynamics along certain directions on the moduli space. 
One such example is the analysis of
\cite{Aharony:1995ne}. These authors considered the dynamics of  an $A_2$ model along the Higgs branch parameterized by the meson $M_1\sim Q\bar Q$.  Choosing the rank of the meson matrix to be $F_M<N-1$, one finds the electric theory gauge group is broken to $SU(N-F_M)$. 
The adjoint $X$ decomposes into an adjoint $X^\prime$, $F_M$ fundamental flavors $Q^\prime$, $\overline Q^{\prime }$ and gauge singlets. 
Since $F_M$ fundamental flavors are eaten by super-Higgs mechanism, the number of light fundamentals is still $F$. 
On the other hand, the global symmetry is broken to $SU(F-F_M)$ due to a superpotential term 
\begin{equation}
	\label{eq:ASYdeformation}
	\Delta W_{el}=\sum_f^{F_M}\overline Q^{\prime f}X^\prime Q^{\prime}_f\,
\end{equation}
which arises from the $X^3$ superpotential of the UV theory. 

A naive duality transformation suggests that on the Higgs branch the magnetic dual is described in terms of an $SU(2F-(N-F))$ theory, i.e. it the gauge group of this dual is larger than the $SU(2F-N)$ magnetic group in the UV. The resolution of this apparent puzzle arises from the observation that  the superpotential deformation (\ref{eq:ASYdeformation}) leads to tadpoles in the superpotential of the magnetic theory, which in turn break the $SU(2N-(F-N))$ group down to the expected $SU(2F-N)$.

A more general analysis of ADE models was performed in \cite{Intriligator:2003mi,Intriligator:2016sgx}. 
While the authors found that the duality conjectures worked in most cases, the duality conjecture failed along some flat directions in even-$k$ $D_{k+2}$ models. The authors of \cite{Intriligator:2016sgx} performed an extensive analysis of the dynamics but did not find non-perturbative effects that would lift the dangerous flat directions and restore the duality conjecture. 
Going beyond $D_{k+2}$ models, it was shown in \cite{Intriligator:2016sgx} that similar puzzles arise in $E_7$ superconformal field theories where different RG flow trajectories resulted in different number of higher dimensional representations in the IR.

\subsection{Dynamics of Nilpotent Flat Directions }
\label{sec:nilpotent}

Despite a large body of work, the dynamics along the flat directions of ADE models has not been studied exhaustively. 
While a complete analysis of the moduli space is beyond the scope of our work, we will  here consider a large new class of flat directions, the nilpotent flat directions along which $X\ne 0$ while $X^k=0$.  These flat directions are parameterized by mesons of the form $M_{m}\sim\overline{Q}X^{m-1} Q$. 
While some of these flat directions were considered in \cite{Aharony:1995ne} in the case of $A_k$ models, they have not been studied in $D_{k+2}$ models.

When we move along a  nilpotent direction, we can analyze the RG flow into the IR in two different ways. In the first approach we will start from the magnetic dual theory in the UV. The moduli space of this theory consists of the Higgs and Coulomb branches, parameterized by dressed mesons and baryons of the magnetic theory, as well as a mesonic branch parametrized by gauge singlet mesons $M_{m}$. Moving along along our chosen nilpotent flat direction on the electric side corresponds to turning on the VEVs of these mesons and moving onto {\it the meson branch of the magnetic theory}. The magnetic gauge group remains unbroken along the meson branch but the magnetic superpotential  is modified.

\begin{figure}[h!]
\centering
\vspace{0.3cm}
\begin{tikzpicture}[
	node distance=1.5cm and 1.4cm,
	box/.style={draw, fill=gray!15, minimum width=4.3cm, minimum height=1.2cm, align=center},
	arrow/.style={-{Latex}, thick},
	every node/.style={font=\normalsize}
	]
	
	\node[box] (A) {UV Electric Theory};
	\node[draw=none, minimum width=0pt, minimum height=0pt, right=of A] (Ddummy2) {};
	\node[draw=none, minimum width=0pt, minimum height=0pt, right=of Ddummy2] (Ddummy3) {};
	\node[draw=none, minimum width=0pt, minimum height=0pt, right=of Ddummy3] (Ddummy4) {};
	\node[box, right=of Ddummy4] (B) { UV Magnetic Theory};
	\node[box, below=of A] (C) {  Electric Theory \\on Higgs Branch};
	\node[box, right=of C] (D) {Magnetic dual of \\ Higgs Branch Theory};

	\node[draw=none, minimum width=0pt, minimum height=0pt, below=of D] (Ddummy) {};
     \node[box, right=of Ddummy] (G) {Magnetic Theory on\\ Meson Branch};

	\draw[arrow,dashed] (A) -- (B);
	\draw[arrow] (A) -- (C);
	\draw[arrow,dashed] (C) -- (D);
    \draw[arrow] (D) -- (G);
	\draw[arrow] (B) -- (G);
	
\end{tikzpicture}
\caption{Flows between electric and magnetic theories under Higgsing and meson-branch deformations. The horizontal dashed lines indicate  duality maps, and the vertical or diagonal lines represent flows from the UV theories to the IR theories. 
}
\label{fig:duality_diagram}

\end{figure}
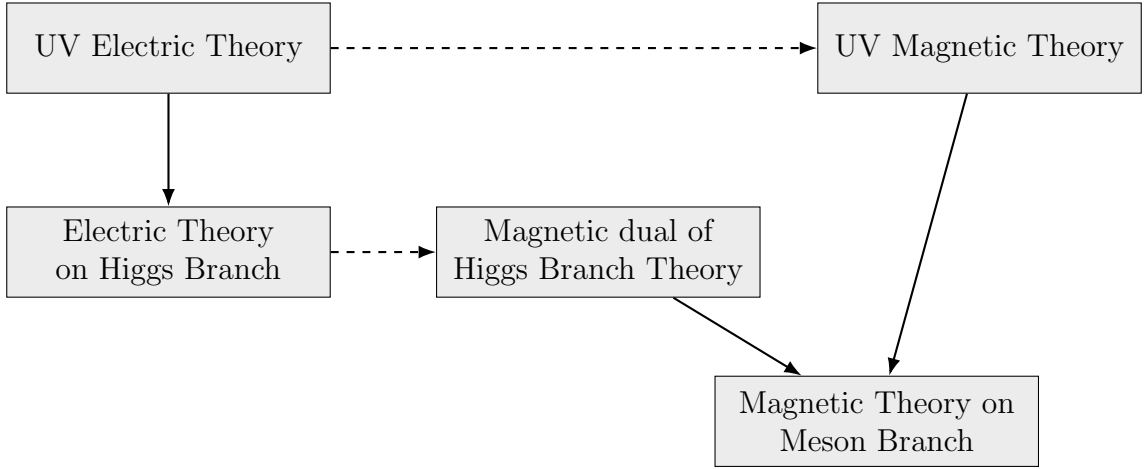

In the second approach we start from the electric theory and  move along the nilpotent flat direction, thus partially breaking the electric gauge group. Integrating out the heavy degrees of freedom we find an effective description in terms of an $A_k$ or $D_{k+2}$ theory with a smaller gauge group and a deformed superpotential. We will refer to this theory as the {\it electric theory on the Higgs branch}. At this point we can perform the duality transformation, thus constructing a {\it magnetic dual of the Higgs branch theory}. Somewhat surprisingly, the gauge group of this magnetic dual is larger than the magnetic gauge group found in the first approach. This is not necessarily a failure of the duality. Indeed, we shall see that the superpotential deformation of the Higgs branch electric theory implies a spontaneous symmetry breaking in its magnetic dual,  leading to a possible agreement with the analysis employed in our first approach. If the duality works the RG flow may be graphically represented as in Figure \ref{fig:duality_diagram}.

Indeed, when studying the $A_k$ models we will show that the theories follow the flows indicated by this diagram, further confirming the validity of the duality conjecture.  On the other hand, we will argue that in $D_{k+2}$ models the RG flow diagram can not close for any value of  $k$, thus invalidating the duality conjecture in this class of models. 

To be more specific, we note that the UV magnetic gauge group in the UV remains unbroken along the mesonic branch. On the other hand, we will find complementary pairs of nilpotent directions of the electric theory along which the effective description will be given in terms of $D_{k+2}$ models with different gauge groups yet the same superpotential deformations. Since, in the magnetic description, the superpotential deformations appear as symmetry breaking tadpoles, this precludes the possibility that the duals of the complementary nilpotent directions can flow to the same gauge group of the UV magnetic dual on the meson branch.
It is interesting that the failure mode will be similar to the one found in \cite{Intriligator:2016sgx}  -- this time different RG trajectories will lead to IR physics with the same superpotential but different magnetic gauge groups.

\subsection{ Duality Tests for $A_k$ Theory}
\label{sec:Aktests}

\subsubsection{The Electric Theory on the Higgs branch}
\label{sec:akmodels}

We  consider a flat direction parameterized by a VEV of a single meson $M_{n}=\overline Q^F \! X^{n-1} Q_F$, $1<n\le k$. In terms of the elementary fields (and up to gauge and global symmetry transformations), this flat direction is given by
\begin{align}
	\label{eq:nilpotentflat}
    X^1_2=X^2_3=\dots=X^{n-1}_{n}=Q_{F}^{n}=\overline{Q}^{F}_1=v\,
\end{align}
where the indices on the fundamental and anti fundamental are color indices. All other components of the fields are 0.  It is easy to see that the adjoint is nilpotent along this direction, $X^{n}=0$.

Along this direction the unbroken gauge group  is $SU(N-n)$ and the adjoint $X$ decomposes into an adjoint $X^\prime$, $n$ fundamental flavors, and several gauge singlets. At the same time, $n$ fundamental flavors are eaten by the super-Higgs mechanism.  Thus the low energy $SU(N-n)$ theory still has $F$ light flavors, one of which is still coupled in the superpotential.. 

The light degrees of freedom and their charges are given by Table \ref{tab:akElIR}
\begin{table}[h!]
\centering
	\def\arraystretch{1.5}
	\begin{tabular}{||c | c | c | c| c| c|} 
		\hline
		Field & $SU(N-n)$  & $SU(F-1)$  & $SU({F-1})$ & $U(1)_B$ & $U(1)_R$\\ [0.5ex] 
		\hline\hline
		$ Q^\prime$ & \fund & \fund & 1&1 &$\frac{F(k+1)+n-2N-1}{(F-1)(k+1)}$\\ 
		\hline
		$\bar{Q^\prime}$ & \antifund  & 1 & \antifund &-1&$\frac{F(k+1)+n-2N-1}{(F-1)(k+1)}$ \\
		\hline
		$x$ & \fund & 1 & 1 &1&$\frac{n+1}{k+1}$\\
		\hline
		$\bar{x}$ & \antifund  & 1 & 1&-1&$\frac{n+1}{k+1}$\\
		\hline
		${X}^\prime$ & Adj & 1 & 1&0&$\frac{2}{k+1}$ \\  
		\hline
	\end{tabular}
\caption{Matter content and global charges of the $A_{k}$ IR electric theory.}
\label{tab:akElIR}
\end{table}
where we omitted gauge singlet fields with the R charge less than $2/3$ since they saturate the unitarity bound and become free, decoupling from the interacting physics at the IR fixed point~\cite{Seiberg:1994pq,Kutasov:2003iy,Mack:1975je,Intriligator:2003jj}.
The electric superpotential of the low energy theory takes the form
\begin{align}
    \mathrm{Tr}(X^{k+1})&= \mathrm{Tr}\left(({X^\prime})^{k+1}\right)+\bar{x} \left({X^\prime}\right)^{k-n}x+\dots
\end{align}
where $x$ and $\overline x$ represent the light linear combination of the moduli left uneaten by the super-Higgs mechanism. 

We note that one can interpret this effective theory as a deformation of an $A_k$ model with $N^\prime=N-n$ colors and $F$ flavors by a superpotential term involving just one fundamental flavor. The deformation breaks the global symmetry to $SU(F-1)$.

\subsubsection{The Magnetic Theory on the meson branch}

The magnetic dual of the $A_k$ theory along the nilpotent direction of the moduli space can be constructed in two ways. First we note that $\left(M_n\right)^F_F$ direction on the Higgs branch of the electric theory corresponds to a gauge singlet direction of the magnetic theory (we will refer to it as a mesonic direction). Below the scale of the meson VEV the physics is still described in terms of the $SU(kF-N)$ magnetic gauge group but now with the superpotential
\begin{align}
	\label{eq:AkUVmag}
   W_{mag}&= \mathrm{Tr}(\tilde{X}^{k+1})+\sum_{m} M_m \bar{q}\tilde{X}^{k-m} q +\bar{q}^{F}\tilde{X}^{k-n} q_{F}\,,
\end{align}
although one should remember that the Yukawa coupling in the last term is given by a VEV of a massless field.

\subsubsection{The Dual of the Electric Theory on Higgs  Branch}

We can now perform the duality transformation on the electric theory on the meson  branch. Following \cite{Aharony:1995ne} we find that this magnetic theory has an $SU(kF-N+n)$ gauge group,  matter content given by Table \ref{tab:akElIRDual}.
\begin{table}[h!]
\centering
\def\arraystretch{1.5}
\begin{tabular}{||c | c | c | c| c| c|} 
 \hline
 Field & $SU(kF-N+n)$  & $SU(F-1)$  & $SU({F-1})$ & $U(1)_B$ & $U(1)_R$\\ [0.5ex] 
 \hline\hline
 $ q$ &\fund & \antifund & 1&$\frac{N-n}{kF-N+n}$& $\frac{F(1-k)+2N-n}{(F-1)(k+1)}$ \\ 
 \hline
 $\bar{q}$ & \antifund  & 1 &\fund&$-\frac{N-n}{kF-N+n}$& $\frac{F(1-k)+2N-n}{(F-1)(k+1)}$  \\
 \hline
 $q_F$ & \fund & 1 & 1 &$\frac{N-n}{kF-N+n}$&$-\frac{n}{k+1}$\\
 \hline
 $\bar{q}^{{F}}$ & \antifund  & 1 & 1 &$-\frac{N-n}{kF-N+n}$&$-\frac{n}{k+1}$  \\
 \hline
 $\tilde{X}$ & Adj & 1 & 1 &0 &$\frac{2}{k+1}$ \\  
 \hline
 $(M^\prime_m)_f^{f^\prime}$&1&\fund&$\antifund$&0&$\frac{2(m(F-1)+Fk +n -2N-1)}{(F-1)(k+1)}$\\
 \hline
 $(M^\prime_m)_F^{f}$&1&1&\antifund&0&$\frac{F(2m+k+n+1)-2(m+N+1)}{(F-1)(k+1)}$\\
 \hline
 $(M^\prime_m)_f^{{F}}$&1&\fund&1&0&$\frac{F(2m+k+n+1)-2(m+N+1)}{(F-1)(k+1)}$\\
 \hline
 $(M^\prime_{m})_F^F$&1&1&1&0&$\frac{2(m+n+1)}{k+1}$\\
 \hline
 $(M^\prime_{k-n})_F^F$&1&1&1&0&2\\
 \hline
\end{tabular}
\caption{Matter content and global charges of the $A_{k}$ Dual of the Electric Theory on Higgs  Branch.}
\label{tab:akElIRDual}
\end{table}
 and the superpotential 
\begin{align}
   W^\prime_{mag}&= \mathrm{Tr}(\tilde{X}^{k+1})+\sum_{m} M^\prime_m \bar{q}\tilde{X}^{k-m} q +(M^\prime_{k-n})^{F}_{F}\,.
\end{align}
To fully analyze the vacuum structure of the deformed magnetic dual, one must observe a presence of a tadpole for the $M^\prime_{k-n}$ meson in the superpotential. This tadpole drives the VEVs of the composite $\bar q \tilde{X}^{ n}q$ meson to non-zero value
\begin{align}
	0&\neq\bar{q}\tilde{X}^{k-(k-n)}q=\bar{q}\tilde{X}^{n}q\,.
\end{align}
At the minimum of the superpotential the unbroken gauge group is $SU(kF-N)$ in an agreement with the 
gauge group of the 
magnetic theory on the meson branch.
 Moreover, integrating out the heavy fields, one arrives at the superpotential (\ref{eq:AkUVmag}).

We therefore find that the duality conjectures work along  these directions in the moduli space of the $A_k$ theory.

\subsection{Duality Tests for $D_{k+2}$ Theory}
\label{sec:Dktests}

We will now proceed in the same manner for the $D_{k+2}$ theory.  However, to analyze the $D_{k+2}$ theory we will consider two { related} flat directions,  parameterized by a  pair of dressed mesons $M_{\tilde n,1}=\overline{Q}^F X^{\tilde n-1} Q_F$ where the first meson has $\tilde n=n$ and the second meson has $\tilde n=k-n$.
It is clear that the gauge symmetry breaking pattern along these flat directions can be inferred from the analysis of the $A_k$ models. Thus we can build on the results obtained in section \ref{sec:akmodels}. 

\subsubsection{The Electric Theory on the Higgs branch} 
Along the $M_{\tilde n,1}$ direction of the Higgs branch the $SU(N)$ gauge group is broken to $SU(N-\tilde n)$ subgroup. Each of the adjoints, $X$ and $Y$, decomposes into an adjoint ($X^\prime$ and $Y^\prime$ respectively), $\tilde n$ new fields in the fundamental representation  ($x, \bar x$, and $y, \bar y $ respectively) and several gauge singlet fields. The low energy superpotential takes the form:
\begin{equation}
\begin{split}
    W &= \mathrm{Tr} \left((X^{\prime})^{ k+1}\right)+\mathrm{Tr} \left({X^\prime (Y^{\prime})^{ 2}}\right)+\bar{x}\left(X^{\prime}\right)^{ k-\tilde n}x + \bar{y}_1Y^\prime x + \bar{x}Y^\prime  y_{\tilde n}
    \\
    &+ \sum_{f=1}^{\tilde n}\bar{y}_fX^\prime y_f+ \sum_{f=1}^{\tilde n-1}\bar{y}_{f+1}y_f+\ldots\,,
\end{split}
\end{equation}
where $x$, $\bar x$, $y_f$, and $\bar y_f$ represent the $2\tilde n$ new flavors in the fundamental representation. It is easy to see that this superpotential gives mass to $\tilde n-1$ flavors while $\tilde n$ flavors are eaten by the super-Higgs mechanism. After integrating out the massive fields, the superpotential becomes
\begin{align}
	 W&= \mathrm{Tr} \left((X^{\prime})^{ k+1}\right)+\mathrm{Tr} \left({X^\prime (Y^{\prime})^{ 2}}\right)+\bar{x}{X^{\prime}}^{ k-\tilde n}x+\bar{y} {Y^\prime}x +\bar{x}{Y^\prime}y+ \bar{y}\left({X^\prime}\right)^{\tilde n}y\,,
\end{align}
where  $\bar{y}=\bar{y}_1$ and ${y}={y}_{n}$ while the matter content of the theory is summarized in Table (\ref{tab:dkElIR}).
\begin{table}[h!]
\centering
\def\arraystretch{1.5}
\begin{tabular}{||c | c | c | c| c| c|} 
 \hline
 Field & $SU(N-n)$  & $SU(F-1)$  & $SU({F-1})$ & $U(1)_B$ & $U(1)_R$\\ [0.5ex] 
 \hline\hline
 $ Q^\prime$ & \fund & \fund & 1 & 1&$\frac{F(k+1)-N-1+n}{(F-1)(k+1)}$ \\ 
 \hline
 $\bar{Q}^\prime$ & \antifund & 1 & \antifund & -1&$\frac{F(k+1)-N-1+n}{(F-1)(k+1)}$\\
 \hline
 $x$ & \fund & 1 & 1& 1 & $\frac{n+1}{k+1}$  \\
 \hline
 $\bar{x}$ & \antifund  & 1 & 1& -1  & $\frac{n+1}{k+1}$  \\
 \hline
  $y$ & \fund &1 & 1& 1 & $1-\frac{n}{k+1}$\\
 \hline
 $\bar{y}$ & \antifund  & 1 & 1& -1& $1-\frac{n}{k+1}$ \\
 \hline
 $X^\prime$ & Adj & 1 & 1 & 0 & $\frac{2}{k+1}$\\  
 \hline
 $Y^\prime$ & Adj & 1 & 1 & 0 & $\frac{k}{k+1}$\\  
 \hline
\end{tabular}
\caption{Matter content and global charges of the $D_{k+2}$ IR electric theory.}
\label{tab:dkElIR}
\end{table}

\subsubsection{The Magnetic Theory on the meson branch}
Starting in the UV the magnetic dual of the $D_{k+2}$ model is given  in terms of an $SU(3kF-N)$ gauge group discussed in  Section \ref{sec:dkplus2}. The nilpotent flat directions of the Higgs branch considered in the previous subsection correspond to the mesonic branch of the magnetic description. While the gauge group of the magnetic description remains unbroken, the meson VEV converts one of the superpotential terms into a tadpole for a dressed magnetic meson
\begin{equation}
	M_{\tilde n,1}\overline{q}\tilde X^{k-\tilde n}\tilde Y^2q\longrightarrow \tilde M_{k-n,3}\,.
\end{equation}
This tadpole drives the magnetic theory to a new IR fixed point.

\subsubsection{The Dual of the Electric Theory on Higgs Branch}
As in the case of the $A_k$ models, the analysis is more involved if we construct a dual of the low energy $SU(N-\tilde n)$ electric description. The magnetic theory has an $SU(N^\prime)$ gauge group with $N^\prime=3k(F+1)-N+\tilde n$. The matter content consists two adjoint supermultiplets, $F+1$ flavors in the fundamental representation and a set of elementary mesons. However, the global non-Abelian symmetry under which the fundamentals and mesons transform is only  $SU(F-1)_L\times SU(F-1)_R$.
The superpotential of the dual description is given by a sum of the usual magnetic superpotential (\ref{eq:Dkplus2magneticW}) of a $D_{k+2}$ and the deformation superpotential
\begin{align}
	\label{eq:dkplus2tadpoles}
    \Delta W'_{mag}&\sim\left(M'_{k-\tilde n,1}\right)^{F}_{F}+\left(M'_{1,2}\right)^{F}_{F+1}+\left(M'_{1,2}\right)^{F+1}_{F}+\left(M'_{\tilde n,1}\right)^{F+1}_{F+1}\,.
\end{align}

\subsubsection{The Failure of the Duality Map}

Let us now compare the { duals of the electric theories on the Higgs brach } 
for the cases $\tilde n=n$ and $\tilde n=k-n$. It is easy to see that the two theories have the same unbroken global symmetry and the same superpotential (\ref{eq:dkplus2tadpoles}). On the other hand, they have different gauge groups (specifically $SU(3k(F+1)-N+n)$ and $SU(3k(F+1)-N+k-n)$). In both cases, the number of colors in this dual description is greater than the number of colors in the UV magnetic theory $N^\prime> 3kF-N$.

Just as in the case of $A_k$ models, the gauge group of these magnetic theories has to be broken down to $SU(3kF-N)$ if duality is to work. Indeed, the tadpoles (\ref{eq:dkplus2tadpoles})  induce VEVs for four dressed magnetic mesons:
\begin{equation}
	\label{eq:dkplustwomagvevs}
\begin{aligned}
	(M^\prime_{k-n,1})_{F}^{F}&\rightarrow \bar{q}_{F} \tilde{X}^n \tilde{Y}^2 q_{F}\neq0\\
	(M^\prime_{1,2})_{F}^{F+1}&\rightarrow \bar{q}_{F} \tilde{X}^{k-1} \tilde{Y} q_{F+1}\neq0\\
	(M^\prime_{1,2})_{F+1}^{F}&\rightarrow \bar{q}_{F+1} \tilde{X}^{k-1} \tilde{Y} q_{F}\neq0\\
	(M^\prime_{n,1})_{F+1}^{F+1}&\rightarrow \bar{q}_{F+1} \tilde{X}^{k-n} \tilde{Y}^2 q_{F+1}\neq0
\end{aligned}
\end{equation}
partially breaking the magnetic $SU(N^\prime)$ group. However, our choice of $\tilde n$ ensures that these VEVs are invariant under the change $n\rightarrow k-n$. It is clear that the same VEVs can not break both $SU(3k(F+1)-N+n)$ and $SU(3k(F+1)-N+k-n)$ down to $SU(3kF-N)$. 
In other words, the duality  must fail for at least one of these nilpotent directions.

\section{Discussion}

Duality conjectures  in N=1 supersymmetric theories have been tested in a variety of ways, including matching of symmetries and 't Hooft anomalies. We have analyzed here a new set of  tests, which  involve analysis of dynamics along nilpotent directions on the moduli space.
The construction of the duality map along such flat directions leads to two potential magnetic duals which must be equivalent in the infrared.

We have applied these tests to $\mathcal{N}=1$ duality  with adjoint superfields and nontrivial superpotentials. We have found that models with  $A_k$ superpotentials theory indeed pass this test, lending further support to the duality conjecture. On the other hand, the models with $D_{k+2}$ theories with two adjoints fail the test, suggesting that the conjecture is invalid. While previous authors had already pointed out difficulties on the duality conjecture for even $k$, we emphasize that our tests show that the conjecture  fails for both $k$ odd and $k$
even. 

There are several further directions to pursue. These tests can be applied to other theories with superpotentials, as for example, theories with symmetric and antisymmetric tensors~\cite{Intriligator:1995ax,Luty:1996cg}. It would also be interesting to study a-maximization in these theories. We leave these and other questions for future work.

\section{Acknowledgments}
This work was supported in part by NSF PHY-2210283.

\bibliographystyle{JHEP}

\end{document}